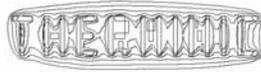

# Evaluation of Cooling Solutions for Outdoor Electronics

Mahendra Wankhede, Vivek Khaire, Dr. Avijit Goswami
Applied Thermal Technologies India
3PrdP Floor, C-Wing, Kapil Towers,
Dr. Ambedkar Road, Pune – 411001
Maharashtra, India
Ph-91-20-66030625
Fax-91-20-66030626
Email: wankhede@appliedthermal.co.in, khaire@appliedthermal.co.in

Prof. S. D. Mahajan
PIET's College of Engineering, Shivajinagar,
Pune-411005, India
Email: sdm@mech.coep.org.in

*Abstract*-The thermal management of an outdoor electronic enclosure can be quite challenging due to the additional thermal load from the sun and the requirement of having an air-sealed enclosure. It is essential to consider the effect of solar heating loads in the design process; otherwise, it can shorten the life expectancy of the electronic product or lead to catastrophic failure. The main objective of this work is to analyze and compare the effectiveness of different cooling techniques used for outdoor electronics. Various cooling techniques were compared like special coats and paints on the outer surface, radiation shield, double-walled enclosure, fans for internal air circulation and air-to-air heat exchangers. A highly simplified, typical outdoor system was selected for this study measuring approximately 300x300x400 mm (WxLxH). Solar radiation was incident on 3 sides of the enclosure. There were 8 equally spaced PCBs inside the enclosure dissipating 12.5W each uniformly (100 watts total). A computational fluid dynamics (CFD) model of the system was built and analyzed. This was followed by building a mock-up of the system and conducting experiments to validate the CFD model. It was found that some of the simplest cooling techniques like white oil paint on the outer surface can significantly reduce the impact of solar loads. Adding internal circulation fans can also be very effective. Using air-to-air heat exchangers was found to be the most effective solution although it is more complex and costly.

*Nomenclature*
PCB           Printed Circuit Board
R              Thermal Resistance, °C/W
T              Temperature, °C
Δ              Difference
α              Solar Absorptivity
ε              Infrared Emissivity

*Subscripts*
amb     ambient
conv    convection
i         inside enclosure
rad      radiation

## I. INTRODUCTION

Outdoor enclosures for housing electronic circuit boards are widely used in variety of technologies including telecommunications, industrial and military applications. These enclosures protect the equipment against a wide variety of environmental hazards, such as sun, moisture, dust and debris. As electronic components have become more powerful and complex, thermal management has become a critical issue. Power dissipation from the equipment can build up over time. In addition, solar load further complicates this problem depending upon the size of the enclosure, surroundings and orientation with respect to the sun [1]. Ignoring the effect of the sun can result in excessively high internal enclosure temperatures causing equipment reliability problems or even failure.

A large variety of cooling techniques have been proposed and used to cool outdoor electronic enclosures. These include conventional techniques, ranging from passive natural convection to the use of commercial air conditioners or heat pumps and concepts using thermosyphons and PCMs (Phase Change Materials) [2]. The internal heat is transferred primarily by convection to the inside surfaces of the enclosure, by conduction through the walls of the enclosure and then by convection to the external ambient. Figure 1 is the simplified representation of an outside enclosure which shows all the thermal resistances that determine the internal temperature $T_i$ of the enclosure [3].

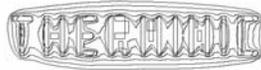

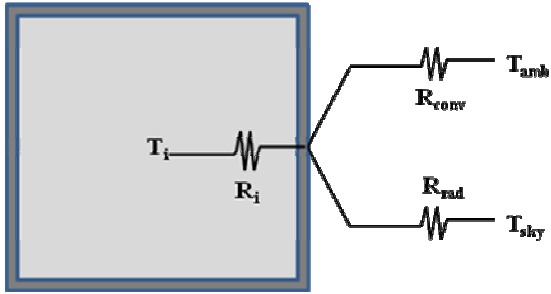

Fig 1: Simplified thermal model of an Outdoor Enclosure

## II CFD SIMULATION CONFIGURATIONS

The CFD simulations were done with "ICEPAK", specialised software for electronics cooling. Total 12 types of cooling configurations were simulated, out of them 9 configurations were created by making combinations of three types of commonly used airflow management options and three types of enclosure surface coatings. Besides these three special configurations were selected which included solar radiation shield, double wall enclosure and a heat exchanger; using a typical aluminium enclosure measuring 300x300x400 mm (WxLxH). There were 8 printed circuit boards (PCBs) inside, each dissipating 12.5W uniformly bringing the total internal power dissipation to 100W. The distance between each PCB was 30mm. The PCB measured 240x180mm and 3mm thick. There was a gap of 35mm between the extreme PCBs and the side faces of the enclosure as shown in fig. 2. Three types of enclosures were selected – first is completely sealed; second sealed with internal circulation fans and the third having vents that allow air exchange with the outside. Three varieties of outside coating were examined including white, black and no coating (plain aluminium finish). These coatings were selected based on their radiation characteristics i.e. solar absorptivity ($\alpha$) and radiation emissivity ($\varepsilon$). White oil coating had a low value of $\alpha$ (0.25) and a high value of $\varepsilon$ (0.91) making it very favourable for cooling under solar heat loads. Black coating had a high value of $\alpha$ (0.88) and a high value for $\varepsilon$ (0.88) while the plain aluminium finish had low values for $\alpha$ and $\varepsilon$ (0.08, 0.09) [4]. In order to minimize the effect of solar loads, various options were analyzed including having a radiation shield, having a double-walled enclosure with air circulation (fig. 3) and a sophisticated air-to-air heat exchanger (fig. 4).

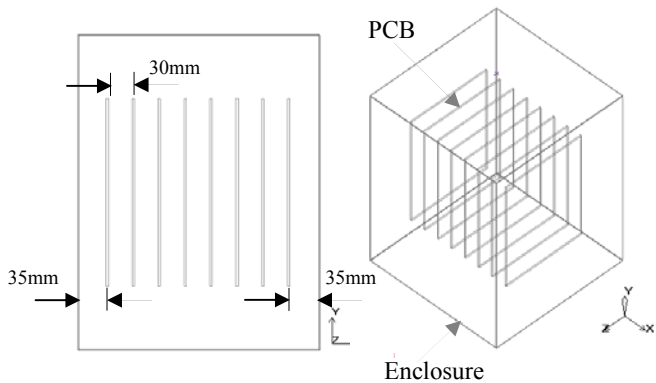

Fig. 2: Icepak Model of Enclosure with PCB assembly

The sealed enclosure with internal fans case was modelled as two fans placed above the PCB assembly and oriented in suction mode pulling the air through the PCB assembly slots. The fan selected was 80x80x20mm and had a maximum flow capacity 48.20 CFM (cubic feet per minute) and a maximum pressure capacity of 6.22mm of H2O. The enclosure with louvered vents case was modelled as two opposite enclosure walls perpendicular to the PCBs having cut-outs at the bottom and top side. The double walled enclosure had an air gap of 20mm between the inner and outer walls. A fan was placed at the top of the outer wall sucking the air between the inner and outer walls through four openings at the bottom of each outer side wall (fig. 3). For this case the fan selected was a 120x120x25mm having a maximum flow capacity 113.11 CFM and a maximum pressure capacity of 10.92mm of H2O. For the case with radiation shield, an umbrella was used to shield the sunlight. In the simulation model, this was modelled as zero solar loads on the enclosure and no radiation heat transfer from the top wall. The heat exchanger system consisted of two parallel heat pipes with condenser ends coming out of the enclosure (fig. 4). Fins were attached to the heat pipes inside the enclosure as well as outside the enclosure. Fans were used to blow air through both sets of fins. Fans selected for this case were 80x80x20mm and had a maximum flow capacity 48.20 CFM and a maximum pressure capacity of 6.22mm of HB2BO. Fig. 4 also shows a schematic of the air circulation pattern and the heat flow from inside the enclosure to the outside.

The following represents a summary of all the configurations analyzed and the abbreviations used.

    NC_SL     : Sealed enclosure without coating,
    NC_LV    : Louvered enclosure without coating,
    NC_WF   : Sealed enclosure without coating and with internal fans,
    BC_SL     : Sealed enclosure with black coating,
    BC_LV    : Louvered enclosure with black coating,
    BC_WF   : Sealed enclosure with black coating and with internal fans,
    WC_SL    : Sealed enclosure with white coating,
    WC_LV    : Louvered enclosure with white coating
    WC_WF: Sealed enclosure with white coating and with internal fans,
    WC_SL_SHLD: White coated sealed enclosure with radiation shield,
    WC_SL_DW: White coated sealed double wall enclosure,
    WC_SL_HX: White coated sealed enclosure with a heat exchanger.

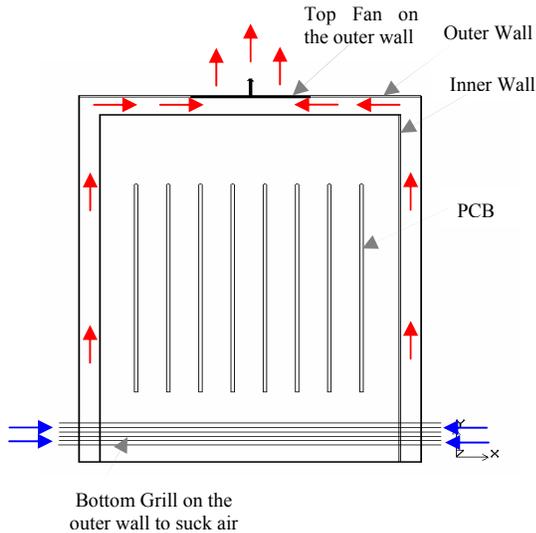

Fig. 3: Schematic of Double wall enclosure

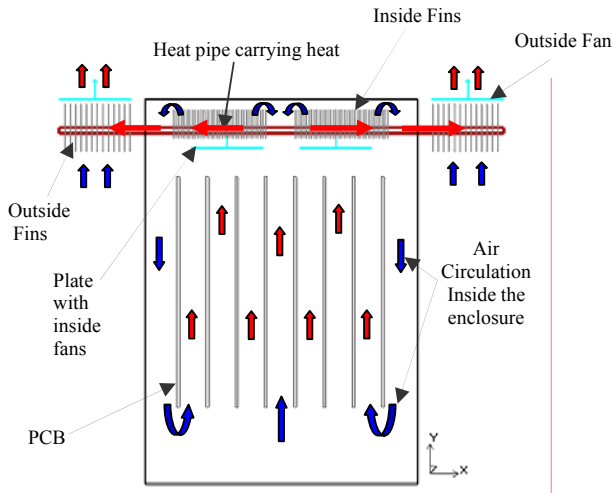

Fig. 4: Heat Transfer mechanism inside enclosure with heat exchanger system

Simulations were run with solar heat load as well as without. For the cases with solar load the maximum load is assumed to be 600 W/m$^2$ for Pune city (India) in the month of March [5] which is one of the hottest months of the year. The sun's rays were assumed to be incident on three adjacent surfaces – the top surface and two sides. This works out to a total solar load of anywhere between 0 and 200W depending upon the solar absorptivity of the outer surface. This shows that the solar load can be of the same magnitude as the internal thermal load.

### III CFD SIMULATION RESULTS WITHOUT SOLAR LOAD

Initially, a set of simulation runs were carried out without any solar heat loads. The following graphs depict the simulation ΔT values for middle PCB surface and the air inside the enclosure (fig. 5a and 5b) without solar load.

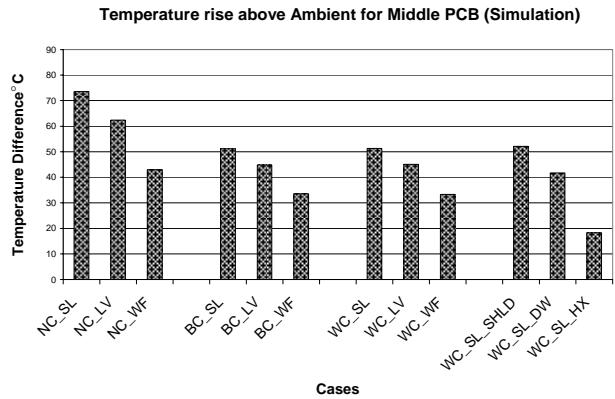

Fig. 5a: Simulation values of ΔT without solar loading for middle PCB

The air temperature inside the enclosure was measured just above the PCB assembly. It can be seen that by just having an internal circulation fan the internal temperatures can be significantly reduced. It is also well below the configuration having vents. In fact, it can be seen that vents have a relatively small impact on the temperatures. Having a black or a white coated surface can also be very effective. The case having a white coated enclosure with heat exchanger had the lowest ΔT values. In this case, a significant amount of heat is dissipated to the outside through the heat exchanger. In the cases of double-walled enclosure and radiation shield there isn't any improvement since there was no solar load.

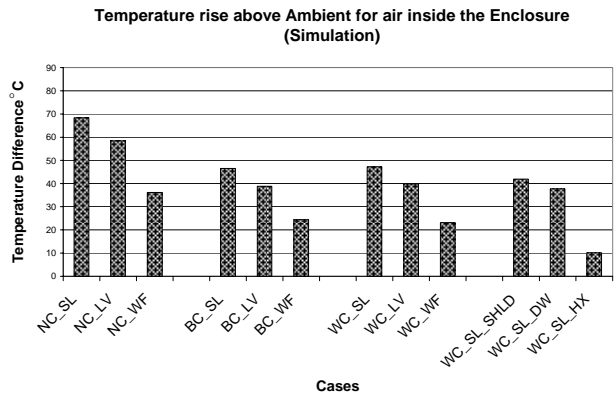

Fig. 5b: Simulation values of ΔT without solar loading for air inside the enclosure

## IV CFD SIMULATION RESULTS WITH SOLAR LOADING

For cases with solar heat load the ΔT values for middle PCB are almost 20% higher for sealed black enclosure (fig. 6a and 6b), showing that solar loading can be substantial for outdoor enclosures. Similar trends were observed compared to the cases without solar loading. The results again show that by just having an internal circulation fan can significantly reduce the internal temperatures. Also, a black or preferably white coated surface can be very effective for cooling compared to a plain aluminium finish. This is because the white coating has a very favourable combination for low solar absorptivity and high radiation emissivity. The temperatures were the least in the case of the air-to-air heat exchanger. It was found that in this case nearly 60 W (40 %) were dissipated through the heat exchanger.

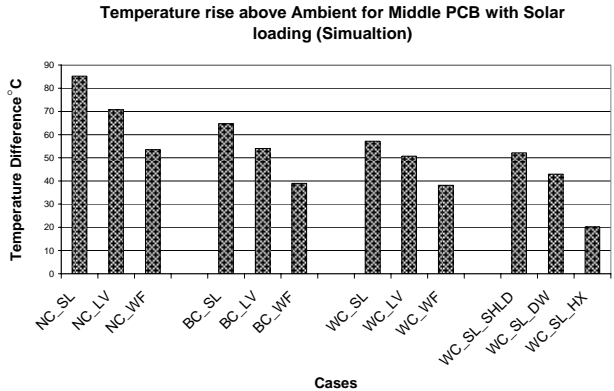

Fig. 6a: Simulation values of ΔT with solar loading for middle PCB

## V EXPERIMENTAL VALIDATION

To validate the CFD results a mock-up of the system was built and tested. Three types of enclosures were built similar to the CFD modelling configurations; perfectly sealed enclosure, one with louvered vents and a sealed enclosure with internal fans. Each configuration was further tested with three types of coats; white coat, black and a plain aluminium finish (fig. 7). Experiments were conducted indoor as well as outdoor conditions. Fig. 8 represents schematic of the enclosure set-up and the temperature measurement locations.

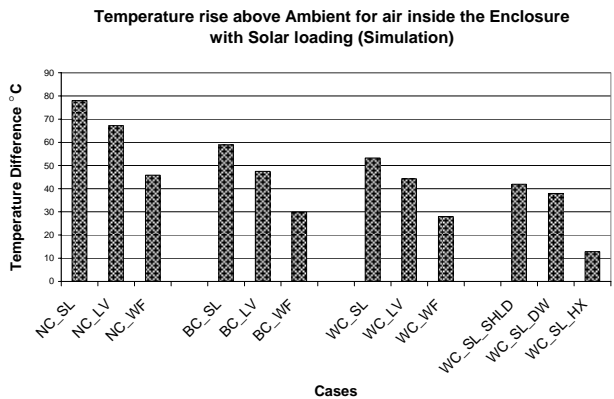

Fig. 6b: Simulation values of ΔT with solar loading for air inside the enclosure

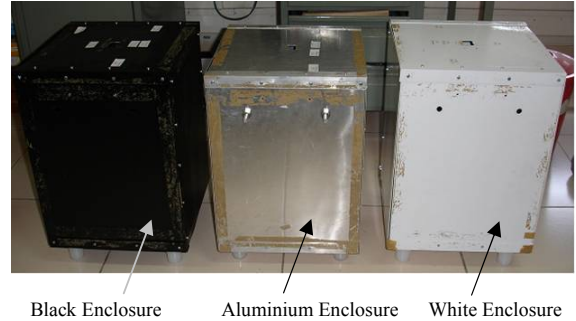

Fig. 7: Enclosures with three different surfaces

The enclosures were constructed using 1mm thick aluminium sheets assembled together using nuts and bolts. The printed circuit boards (PCBs) were made using resistor coils sandwiched between two commercially available copper clad boards (fig. 9). The PCB assembly was placed inside the enclosure suspended from two rods (fig. 10). The thermocouples and DC source wiring was taken out through an opening created on the top side of the enclosure. Each PCB resistors were connected as shown in fig. 11. The power sealed using tapes. After the system was powered using the DC supplied to the enclosure was 100 watts through an external DC power supply and was maintained constant for all the cases. To monitor the temperatures a data acquisition system (DAS) was used. Temperatures were measured on the surfaces of the middle and extreme PCBs and also the enclosure air temperatures were measured at the bottom and at the top of the PCB assembly using T-type thermocouples as shown in fig. 8. For the test case with internal fans, the fans were placed above the PCBs and oriented in the suction mode pulling air through the PCB slots (fig. 12). Fig. 13 shows white coated louvered enclosure with louvered cut-outs at the top and bottom sides.

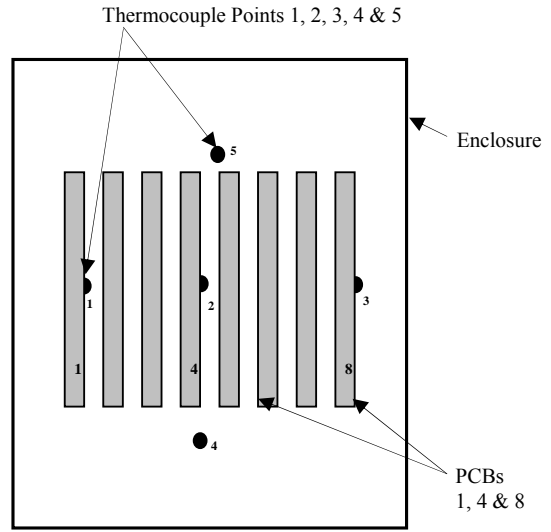

Fig. 8: Schematic of Enclosure Set-up and temperature measurement locations.

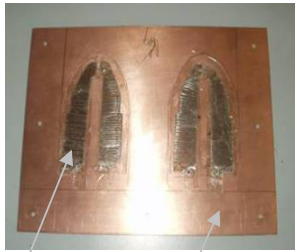

Fig. 9: Resistance coils sandwiched between PCBs

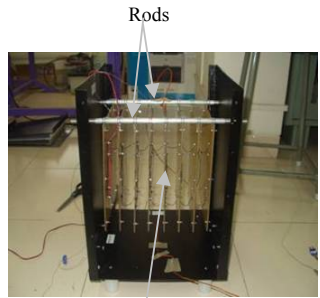

Fig. 10: Arrangement for supporting PCB assembly

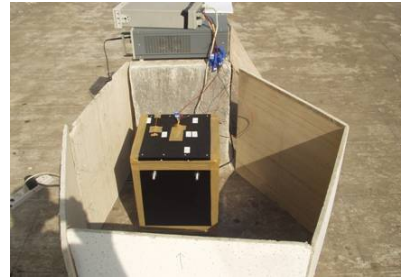

Fig. 14: Experimental set-up for test with solar loading

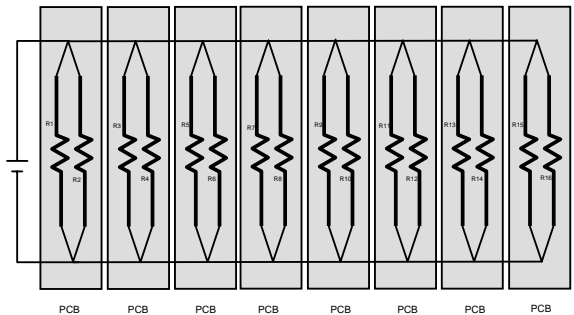

Fig.11: Circuit diagram showing the layout of resistor heaters on the PCBs.

To ensure the system to be air tight all the side gaps were source, it was found that there were significant $I^2R$ losses in the wiring from DC source to the PCB assembly. To ensure 100 watts heat dissipation in the enclosure, the input power from the DC source was increased in such a way as to compensate for the $I^2R$ losses. It was also found that the individual PCB resistances differed slightly. Therefore, the PCBs that were chosen for measurement were only those which dissipated close to 12.5 watts. For the test case with solar loading the black enclosure was tested outside in the sun. Cardboards were used around the enclosure to block off the wind but without blocking the sunlight falling over the enclosure (fig. 14). Temperature readings were recorded after the system reached a steady state which took around couple of hours.

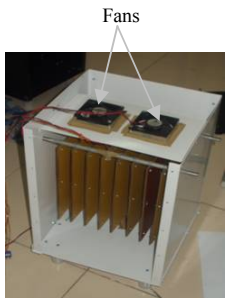

Fig. 12: Enclosure with fans (Top and one side panel removed)

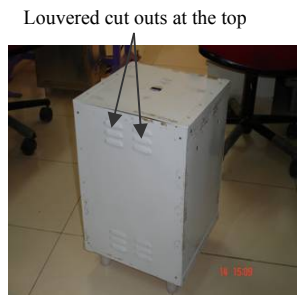

Fig. 13: Louvered Enclosure

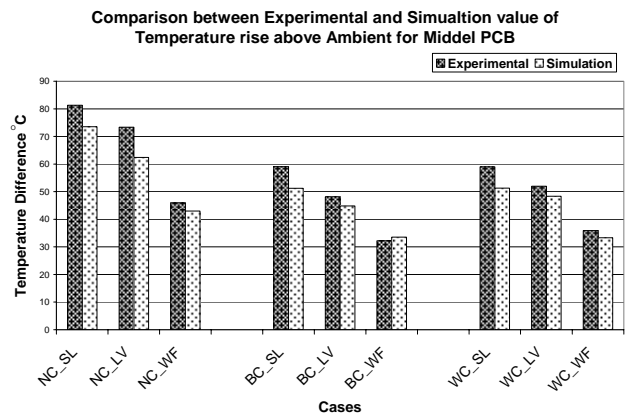

Fig. 15a: Comparison of Simulation and Experimental ΔT value for middle PCB

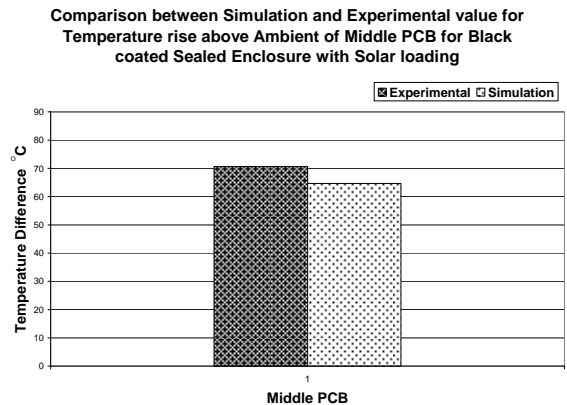

Fig. 15b: Comparison of Simulation and Experimental ΔT value for middle PCB with solar loading for black coated sealed enclosure

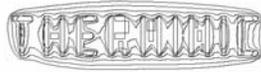

### VI COMPARISON BETWEEN CFD AND TEST RESULTS

From fig. 15a it is seen that the tested ΔT values are quite close to the CFD results for all the tested cases. The percentage difference between the CFD simulation and test results is defined as;

$$\%\Delta = \frac{(Expt.value - simulation.value)}{Expt.value} \times 100$$

It is observed that the difference is within ±10% (fig. 15a) for all the cases with fans. In the other cases the difference is around ±15%. Since in these non-fan cases radiation heat transfer plays a more significant role as compared to the cases with internal fans, the higher differences are probably due to the assumed values of radiation emissivitiy in the simulations.

### CONCLUSION

It was found from this study that the effect of solar heat load on an outdoor system can be quite significant and can increase the internal air temperature by 20%. Different cooling approaches for outdoor electronic enclosures were analyzed and compared. The results indicate the relative effectiveness of these different cooling solutions. Relative to the sealed enclosure without any coating it was found that black or preferably a white coating on the outside enclosure wall is a very simple and effective cooling solution and can reduce internal temperatures by around 25%. It was also found that having vents did not reduce the temperatures as significantly as having internal circulation fans. In fact, there is around 50-55% reduction in the ΔT due to the internal fans compared to a sealed enclosure with no fans. Having a radiation shield and a double-walled enclosure with air circulation provided relatively modest improvements of around 25%. The most dramatic improvement was seen in case of the air-to-air heat exchanger of almost 75%.

### ACKNOWLEDGMENT

Authors of the paper are thankful to Dr. Sukhvinder S. Kang for his valuable review of the paper and suggestion during the work.